\documentclass[twocolumn]{aastex63}
\def\actaa{Acta Astronomica}

\usepackage{amsmath}

\hypersetup{linkcolor=red,citecolor=blue,filecolor=green,urlcolor=magenta}

\begin{document}

\shorttitle{Mira PL relations}
\shortauthors{Ngeow et al.}

\title{Zwicky Transient Facility and Globular Clusters: the $gr$-Band Period-Luminosity Relations for Mira Variables at Maximum Light and Their Applications to Local Galaxies}

\correspondingauthor{C.-C. Ngeow}
\email{cngeow@astro.ncu.edu.tw}

\author[0000-0001-8771-7554]{Chow-Choong Ngeow}
\affil{Graduate Institute of Astronomy, National Central University, 300 Jhongda Road, 32001 Jhongli, Taiwan}

\author[0000-0002-6928-2240]{Jia-Yu Ou}
\affil{Graduate Institute of Astronomy, National Central University, 300 Jhongda Road, 32001 Jhongli, Taiwan}

\author[0000-0001-6147-3360]{Anupam Bhardwaj}
\affil{INAF-Osservatorio astronomico di Capodimonte, Via Moiariello 16, 80131 Napoli, Italy}

\author[0000-0003-1227-3738]{Josiah Purdum}
\affiliation{Caltech Optical Observatories, California Institute of Technology, Pasadena, CA 91125, USA} 

\author[0000-0001-7648-4142]{Ben Rusholme}
\affiliation{IPAC, California Institute of Technology, 1200 E. California Blvd, Pasadena, CA 91125, USA}

\author[0000-0002-9998-6732]{Avery Wold}
\affiliation{IPAC, California Institute of Technology, 1200 E. California Blvd, Pasadena, CA 91125, USA}

\begin{abstract}

  Based on 14 Miras located in 7 globular clusters, we derived the first $gr$-band period-luminosity (PL) at maximum light for the large-amplitude Mira variables using the multi-year light-curve data collected from the Zwicky Transient Facility (ZTF). Since Miras are red variables, we applied a color-term correction to subsets of ZTF light curves, and found that such corrections do not have a large impact on period determinations. We applied our derived PL relations to the known extragalactic Miras in five local galaxies (Sextans, Leo~I, Leo~II, NGC6822 and IC1613), and determined their Mira-based distances. We demonstrated that our PL relations can be applied to short-period ($\lesssim 300$~days) Miras, including those in the two most distant galaxies (NGC6822 and IC1613) in our sample even when only a portion of the light-curves around maximum light have detections. We have also shown that the long-period extragalactic Miras do not follow the PL relations extrapolated to longer periods. Hence, our derived PL relations are only applicable to the short-period Miras, which will be discovered in abundance in local galaxies within the era of Vera C. Rubin Observatory's Legacy Survey of Space and Time.

\end{abstract}


\section{Introduction}\label{sec1}

Mira variables \citep[hereafter Miras, see][for a general review]{mattei1997} are pulsating asymptotic giant branch (AGB) stars with periods longer than 100~days. Miras obey a period-luminosity (PL) relation, especially in the $K$-band or the bolometric magnitudes, because Miras are cool supergiants with radiation peaks in the near-infrared (NIR). Indeed, the majority of the PL relations derived, or calibrated (that is, calibrating the zero-point of the PL relation by fixing the slope), in the literature were in the $JHK$-band (mainly in the $K$-band) and/or being converted to bolometric magnitudes \citep[for examples, see][]{glass1981,glass1982,feast1984,reid1988,feast1989,hughes1990,groe1996,bedding1998,whitelock2000,glass2003,rejkuba2004,soszynski2005,feast2006,whitelock2008,matsunaga2009,tabur2010,ita2011,yuan2017,yuan2018,bhardwaj2019,grady2019,ita2021,andrian2022,sun2022,sanders2023}. Some of these studies also included the derivation of period-Wesenheit (PW) relations, or the addition of a color-term for a period-luminosity-color (PLC) relation. At wavelengths shorter than $J$-band, PL (and PLC) relations have been derived in various optical bands \citep[including $I$-band and Gaia bands,][]{ita2011,bhardwaj2019} in addition to the NIR bands, as well as a single-band PL relation in photographic $m_{pg}$-band \citep{vdb1984} and $I$-band \citep{ou2022}. Beyond $K$-band, several works have also derived the mid-infrared PL relations \citep{glass2009,matsunaga2009,riebel2010,ita2011,iwanek2021}. Finally, based on a comprehensive multi-bands analysis for Miras in the Large Magellanic Cloud (LMC), \citet{iwanek2021b} presented a set of synthetic PL relations in 42 bands ranged from $0.37\mu$m to $25.5\mu$m.

In light of the recent ``Hubble Tension'' \citep[for general reviews, see][]{verde2019,div2021,freedman2021,shah2021,riess2022}, Miras offer an independent route to determine the $H_0$ via the local distance scale ladder, with several advantages being mentioned in \citet{whitelock2013}, \citet{macri2017}, \citet{huang2018,huang2020}, and \citet{sanders2023}. On the other hand, Miras are long-period variables implying the long-term monitoring of extragalactic Miras using space-based telescopes, such as the {\it James Webb Space Telescope (JWST)}, might not be trivial (in terms of scheduling, proposals competitions, etc). On the ground, the 10 years survey of the Vera C. Rubin Observatory's Legacy Survey of Space and Time \citep[LSST,][]{lsst2019} will naturally provide such a long baseline suitable for Miras. Hence, in combination with its photometric depths, LSST is expected to detect numerous Miras in various types of nearby galaxies \citep{macri2017}. Indeed, Miras are included in the Roadmap for the LSST Transients and Variable Stars group \citep{tvs2022} as one of the scientific targets/goals of the survey.

Our aim of this work is to derive the optical band PL relations at maximum light by using the homogeneous light curve data collected from the Zwicky Transient Facility \citep[ZTF,][]{bellm2019,gra19,dec20} for Miras located in the globular clusters. Our work would be similar to the work of \citet{menzies1985} and \citet{feast2002}, but using the most up-to-date and homogeneous globular cluster distances as given in \citet{baumgardt2021}. We choose to derive the PL relations at maximum light because their dispersions were found to be smaller than the mean light counterparts \citep{kanbur1997,bhardwaj2019,ou2022,Ou2023}. Though not an optimal choice, the derived optical band PL relations will still be valuable in the era of LSST, because the six filters set employed by LSST ($ugrizy$) do not extend to the NIR. In this manuscript, Section \ref{sec2} describes the sample of Miras in globular clusters and their extracted ZTF light curves. We then refined their periods in Section \ref{sec3}, followed by the derivation of PL relations at maximum light in Section \ref{sec4}. We applied our derived PL relations to a number of local group galaxies in Section \ref{sec5}. We concluded our work in Section \ref{sec6} together with a brief discussion.

\section{Sample and ZTF Light Curves} \label{sec2}

We started the compilation of Miras in globular clusters from the ``Updated Catalog of Variable Stars in Globular Clusters'' \citep[][hereafter the Clement's Catalog]{clement2001,clement2017}, and only selected variable stars of ``M'' type. V1 and V4 in Palomar~4 was marked as ``SR'' (semi-regular) in Clement's Catalog, however they were reclassified as Mira in \citet{grady2019}. Hence, they were added to the compilation. After excluding known or suspected foreground objects, and V7 in Terzan~5 \citep[because this Mira was found to be a non-member based on radial velocity measurement, see][]{origlia2019}, there were 23 Miras located in 10 globular clusters observable with ZTF ($\delta_{J2000} > -30^\circ$).

\begin{figure}
  \epsscale{1.1}
  \plotone{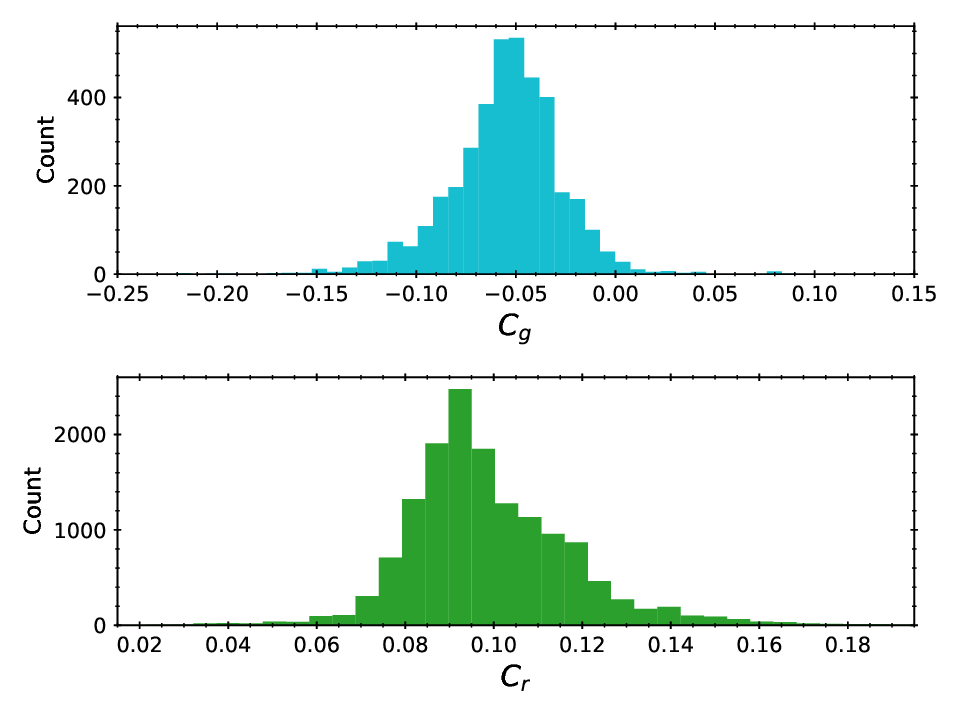}
  \caption{Distributions of color-coefficients, $C_m$, retrieved from the PSF catalogs for our sample of Miras. Upper and lower panels are for the $g$- and $r$-band, respectively.}
  \label{fig_c}
\end{figure}

For this initial sample of globular clusters Miras, we extracted the ZTF $gri$-band time-series photometric data from the PSF (point-spread function) catalogs, produced from a dedicated reduction pipeline \citep{mas19}, using a $1\arcsec$ search radius. The PSF catalogs included those available from the ZTF Data Release (DR) 16 and the ZTF partner survey data\footnote{ZTF observations were divided into three parts, one of them being the partnership survey \citep[for a further details, see][]{bellm2019}.} collected until 2023 March 31. We first visually inspected the ZTF light curves and remove those Miras with small number of data-points, or the data-points only sampling a small portion of the light curves within a single pulsating cycle. This step removed V4 in NGC~6553, V12 and V13 in NGC~6638, V5 and V6 in Terzan~5, V1 in Terzan~12. Three Miras in Terzan~5 either lack $g$-band or only contain very few $g$-band data points, and were subsequently removed because their single $r$-band ZTF light curves cannot be used to estimate colors (see next section). Hence, there are 14 Miras left in 7 globular clusters in the sample.

We have also excluded the $i$-band light curves, because there are 7 Miras in our sample that do not have any ZTF $i$-band data. For the remaining Miras, the numbers of data-points for the $i$-band light curves are in general smaller than other two filters,\footnote{After excluding light curves with null data, the averaged numbers of data points per light curves in the $gri$-band are 242.8, 821.3, and 193.6, respectively.} and many of them only cover a portion of the light curve in a single pulsation cycle, preventing them to be analyzed further. 

\section{Color-Correction on Light Curves and Periods Refinement} \label{sec3}

\begin{figure*}
  \epsscale{1.1}
  \plottwo{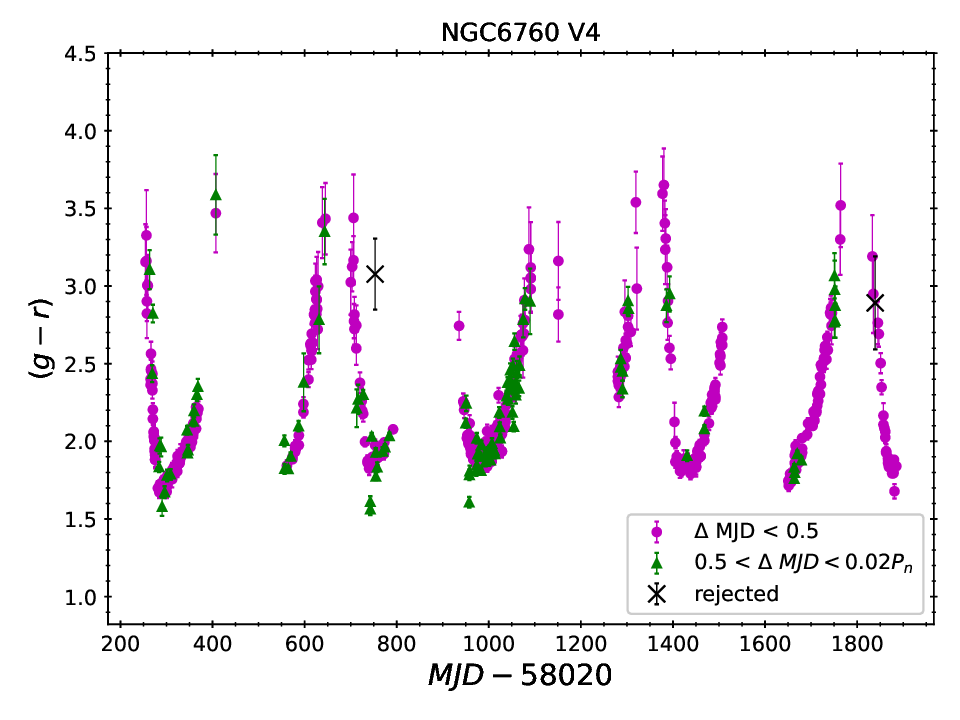}{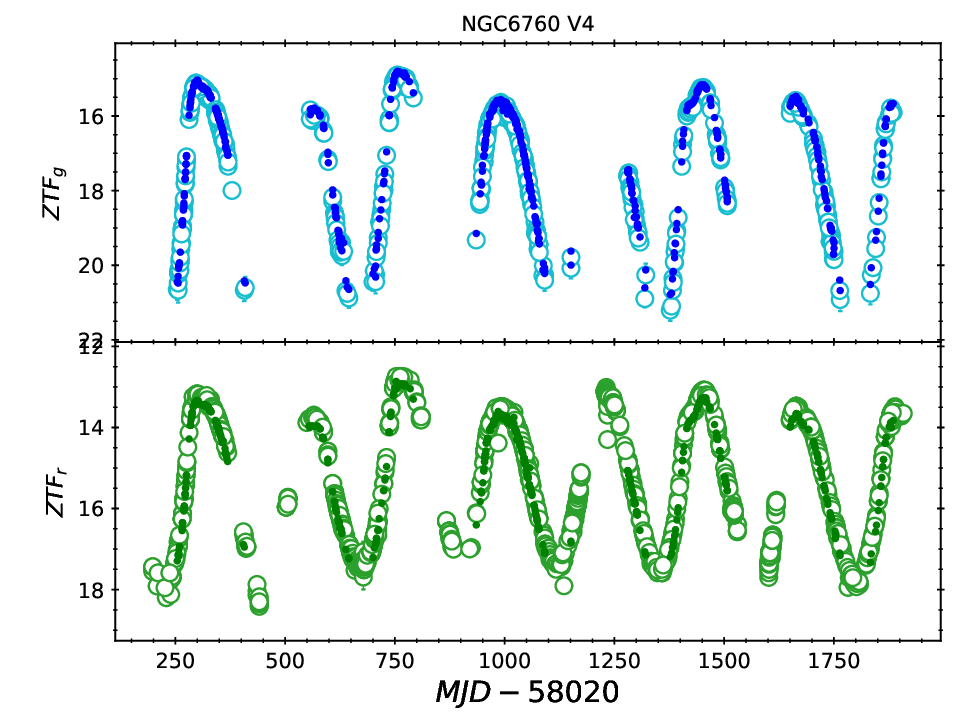}
  \caption{{\bf Left Panel:} Color curve for one of the Mira in our sample constructed using equation (1). The ``MJD'' (modified Julian date) for the $(g-r)$ data-points are at the mid-point of the $gr$-band MJDs. The magenta circles and green triangles are for the pairs of $gr$-band data-points taken in the same night ($\Delta MJD < 0.5$~days) and taken on different nights up to $0.02P_n$~days, respectively. Crosses are rejected data-points either with errors larger than 0.35 or with anomaly $C_m$ values. {\bf Right Panel:} The $gr$-band light curves for the same Miras as in the left panel. Open and filled circles in the $gr$-band light curves are those without and with the color-term corrections, respectively. Note that the color-term corrections did not apply to all $gr$-band data-points, only for those with the $(g-r)$ counterparts. For clarity, error-bars are omitted in the right panels.}
  \label{fig_lcc}
\end{figure*}

As described in \citet{mas19}, PSF photometry in ZTF catalogs were calibrated via $m=m_{ZTF} + ZP_m + C_m (g-r)$, where $m=\{g,\ r\}$ is the calibrated magnitudes in the Pan-STARRS1 \citep{chambers2016,magnier2020} AB magnitude system, and $m_{ZTF}$ represents the ZTF instrumental magnitudes. For non-varying sources, the $(g-r)$ colors can be obtained from the Pan-STARRS1 catalog. For variable stars such as Miras, however, the time-dependent colors have to be known a priori. This is because the $gr$-band observations in ZTF are not simultaneous or near-simultaneous \citep{bellm2019b}. Figure \ref{fig_c} shows the histograms for the color-coefficient, $C_m$, with medians of $-0.053$ and $0.096$ in the $gr$-band, respectively (the corresponding modes are $-0.052$ and $0.091$). Since Miras are very red variable stars, the calibrated light curves should include the $C_m (g-r)$ color-terms. 

Given the long period nature of Miras, the $gr$-band photometry taken within the same night, or within $\sim 0.02P$ (where $P$ is the pulsation period), can be treated as ``near-simultaneous''. Hence, the time-dependent colors can be obtained via the following equation \citep{ngeow2022}:

\begin{eqnarray}
  (g-r) & = & \frac{(g_{ZTF} + ZP_g) - (r_{ZTF}+ZP_r)}{1-C_g+C_r}.
\end{eqnarray}

\noindent We paired up the $gr$-band data-points that are closest in time, up to a threshold of $\Delta MJD < 0.02P_n$ (where $MJD$ is the modified Julian date), to construct the $(g-r)$ colors. The period $P_n$ was determined using the {\tt LombScargleMultiband} module \citep{vdp2015}, available from the {\tt astroML/gatspy} package,\footnote{\url{https://github.com/astroML/gatspy}.} on all of the $gr$-band light curves without the color-terms (hence the subscript $_n$ means no color-terms). Left panel of Figure \ref{fig_lcc} shows an example of the color-curve constructed using equation (1), while the right panels of Figure \ref{fig_lcc} present the $gr$-band light curves without (open circles) and with (filled circles) the inclusion of the color-terms. In general, the pairs of $gr$-band data-points separated by more than a night (green triangles in the left panel of Figure \ref{fig_lcc}) closely resemble those taken within the same night, justifying our assumption that the colors can be constructed for Miras even if the data-points in two bands were taken from different nights.

\begin{figure}
  \epsscale{1.1}
  \plotone{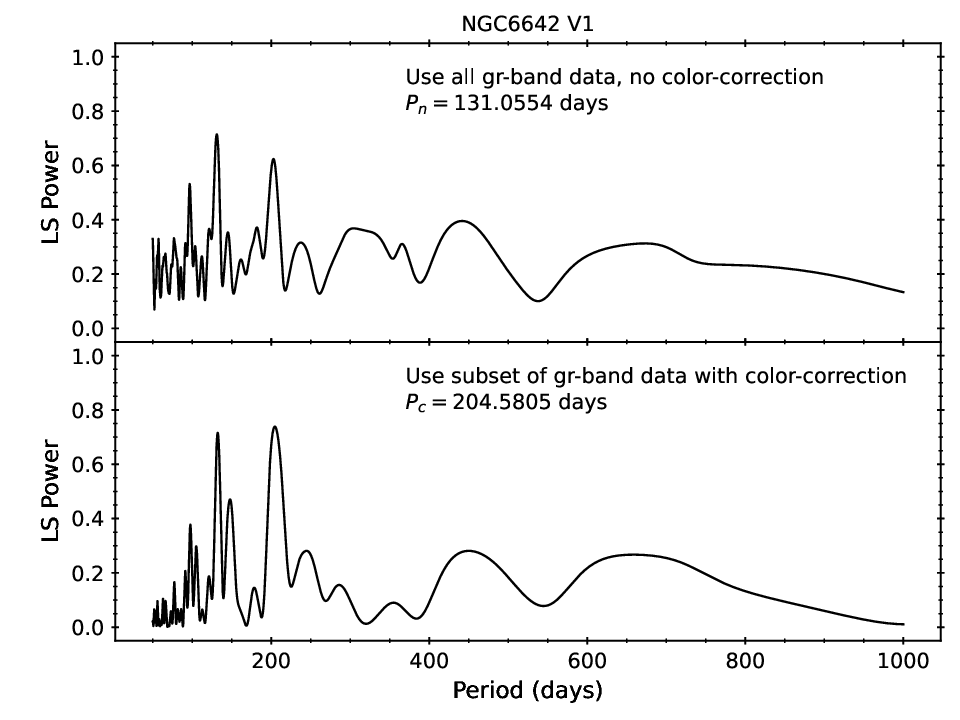}
  \caption{Comparison of the multi-band Lomb-Scargle (LS) periodograms for NGC6642~V1 based on the light curves with (upper panel) and without (lower panel) the corrections of color-terms.}
  \label{fig_ngc6642v1}
\end{figure}

\begin{figure*}
  \centering
  \begin{tabular}{ccc}
    \includegraphics[width=0.32\textwidth]{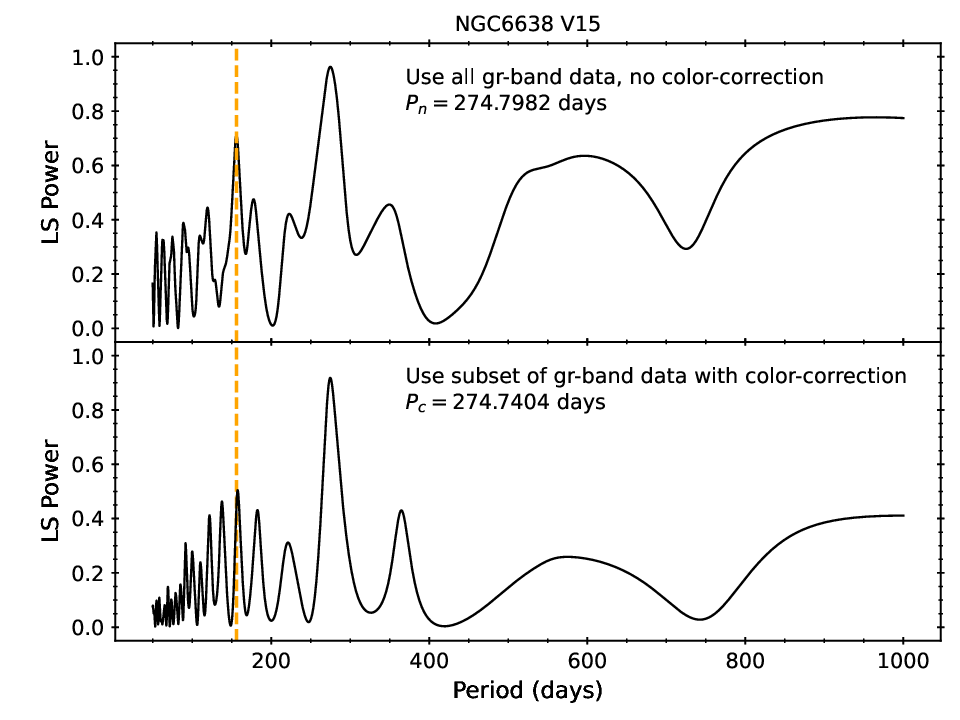} &  \includegraphics[width=0.32\textwidth]{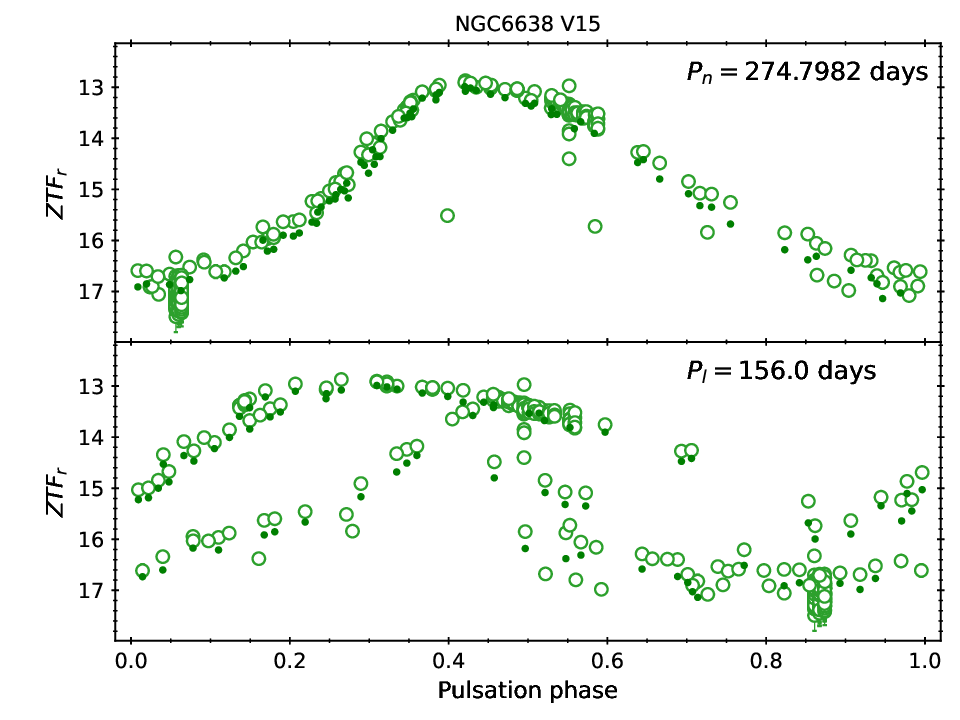} & \includegraphics[width=0.32\textwidth]{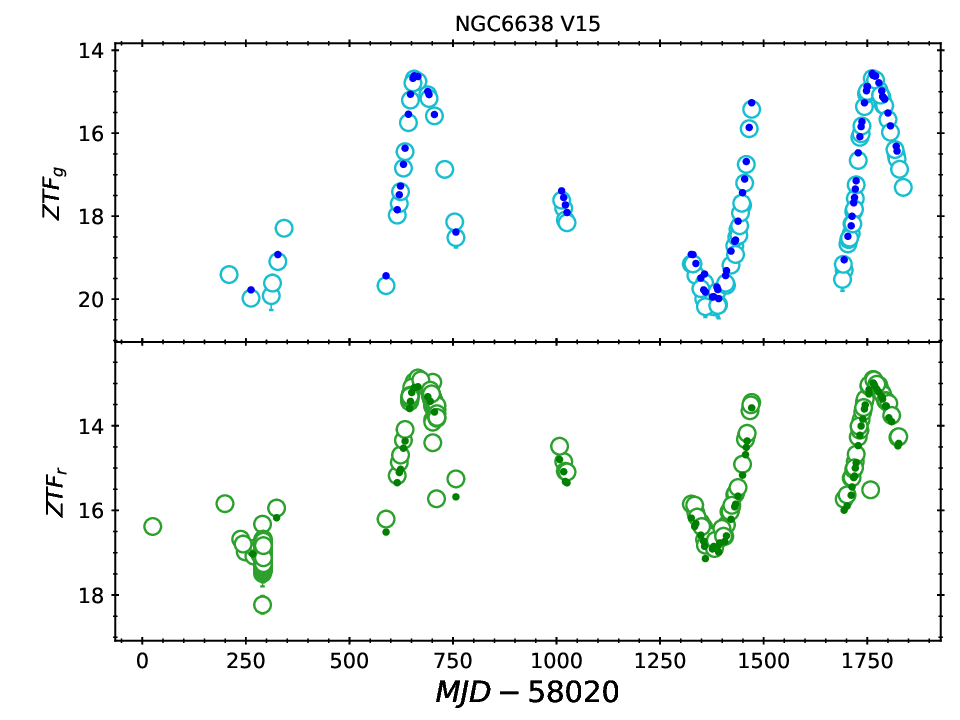} \\
  \end{tabular}
\caption{{\bf Left Panel:} The multi-band LS periodograms based on the ZTF light curves, at which the highest peaks corresponding to $P_n$ (upper panel) and $P_c$ (lower panel). The vertical dashed line indicates the literature period of $P_l=156$~days. {\bf Middle Panel:} Folded $r$-band light curves with our determined period $P_n$ (upper panel) and literature period $P_l$ (lower panel). The filled and open circles are data-points with and without color-terms correction, respectively. {\bf Right Panel:} ZTF light curves and color curve for NGC6638~V15, where the symbols are same as the right panels of Figure \ref{fig_lcc}.}
\label{fig_6638v15}
\end{figure*}

We have also determined the periods based on the subsets of $gr$-band light curve data-points that can be used to calculate the color-terms. When comparing the periods found in these subsets of light curves, the largest percentage difference is $0.1\%$ between the light curves with and without the color-terms correction, suggesting the color-terms do not have a great impact on the determined periods (due to large amplitudes of Miras). The periods based on the subsets of light curves corrected for the color-terms are denoted as $P_c$. In general, $P_c$ is also in good agreement with $P_n$, and the percentage differences between them are less than 1\%. The only exception is NGC6642~V1, for which the full set of $gr$-band light curves, uncorrected for color-terms, give $P_n=131.1$~days. In contrast, the subset of color-corrected light curves returned a period of $P_c=204.6$~days. Figure \ref{fig_ngc6642v1} shows that the Lomb-Scargle (LS) periodograms for this Mira have similar peaks at both periods, suggesting aliasing is affecting its light curves. We adopted $P_n=131.1$~days for NGC6642~V1 because it is closer to the literature period of 127~days. We have also adopted $P_n$ as the final periods for other Miras in our sample, because the total number of data-points for full set of $gr$-band light curves are $\sim2$ to $\sim5$ times more than the subsets of data-points for determining colors, and have a longer time-span (hence, reducing the impact of aliasing). Since $P_n$ and $P_c$ differ by less than 1\%, errors on the adopted $P_n$ are most likely at the 1\%-level.

Finally, with the exception of NGC6638~V15, the percentage differences between our determined $P_n$ and the literature periods ($P_l$) varied from $0.04\%$ to $6.28\%$. In all cases, light curves folded with $P_n$ exhibit a smaller scatter, justifying our determined periods are robust. For NGC6638~V15, the literature period of $P_l=156$~days corresponds to the second highest peaks of the LS periodograms (see left panels of Figure \ref{fig_6638v15}). Nevertheless, our periods of $P_n=274.8$~days can fold the ZTF light curves better than the literature periods, as demonstrated in the middle panel of Figure \ref{fig_6638v15}. For completeness, the right panels of Figure \ref{fig_6638v15} present the $gr$-band light curves and the $(g-r)$ color curve for this Mira.

\section{The PL Relations at Maximum Light} \label{sec4}

\begin{figure*}
  \epsscale{1.1}
  \plottwo{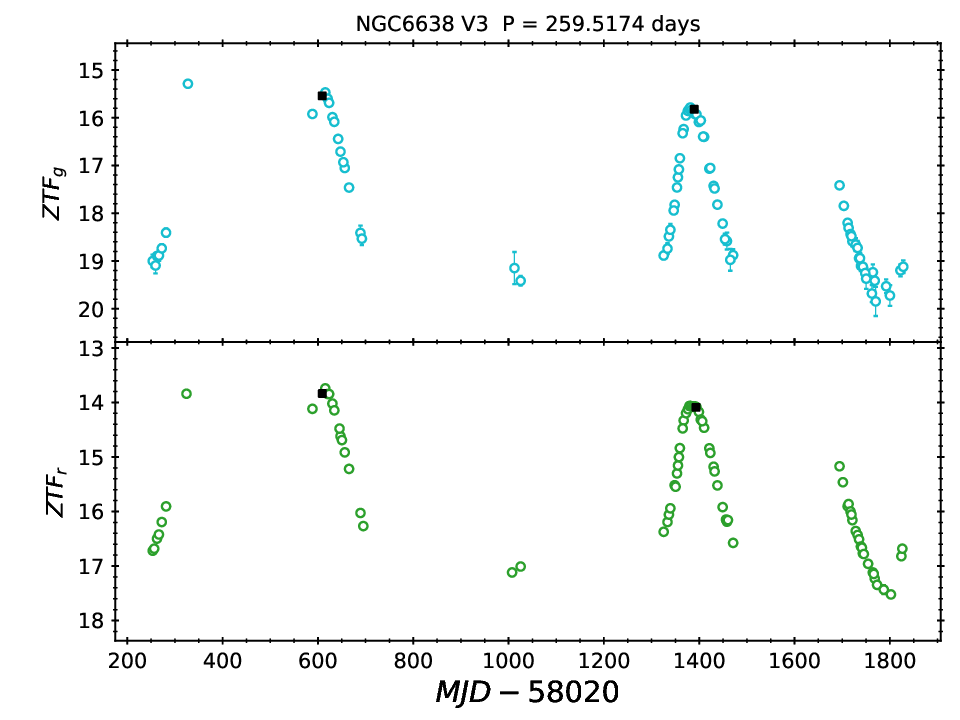}{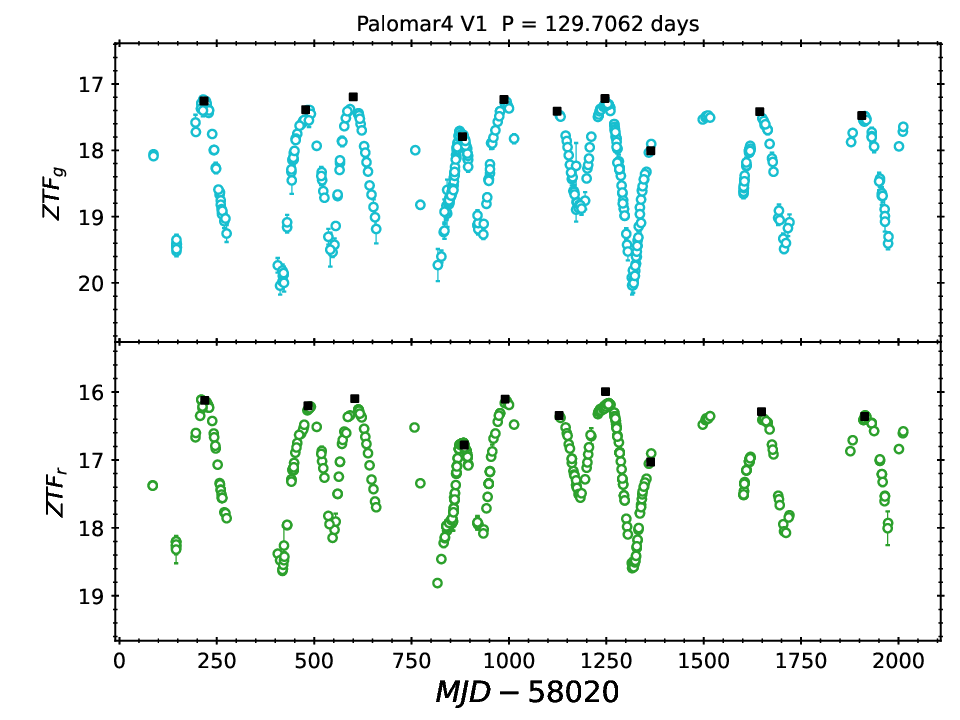}
  \caption{ZTF light curves for two Miras after corrected for the color-terms (open circles). The filled black squares are the determined magnitudes at maximum light, using the method described further in the text. Left panels show a Mira that only two $m^i_X$ can be determined, while the right panels are for a Mira with a total of 10 determined $m^i_X$.}
  \label{fig_magmax}
\end{figure*}

Light curves of Miras are well-known to exhibit cycle-to-cycle variations \citep[for some exemplar light curves, see][]{iwanek2022}. Therefore, \citet{bhardwaj2019} took the averages of the brightest 10\% of the light curves as an estimate of magnitudes at maximum light, $m_X$. While \citet{ou2022} divided the light curves to different pulsating cycles, and took an averaged for $m_X$ found in the pulsating cycles. We followed the approach laid out in \citet{ou2022} to determine the magnitudes at maximum light for our sample of globular cluster Miras. Briefly, we divided the ZTF $gr$-band light curves to a number of pulsation cycles according to the $P_n$ determined in previous section. For $i^{\mathrm{th}}$ pulsation cycle, we fit a sinusoidal function to determine $m_X^i$ if there are more than 10 data-points within the pulsation cycle and contain data-points around the maximum light (see Figure \ref{fig_magmax} for two examples). In general, there are 2 to 10 pulsating cycles which satisfied our selection criteria for determining $m_X^i$, hence we took a mean as the final adopted $m_X$ for our sample of Miras. The standard deviations on $m_X$, $\sigma_X$, were calculated based on small-number statistics \citep[][p. 202]{dean1951,keeping1962}. The values of $m_X$ and $\sigma_X$, together with periods determined in previous section, are summarized in Table \ref{tab1}. In general, $\sigma_X$ are in the order of $\sim 0.5$~mag. This in turns reflect the fact that light curves for Miras are varying from cycle-to-cycle, resulting the magnitudes at maximum light fluctuate at the $\sim 0.5$~mag level.\footnote{In general, amplitudes for shorter wavelengths light curves (such as $gr$-band) would be larger than the $I$-band or $JHK$-band light curves \citep[][their Figure 2]{iwanek2021b}. Hence, it is expected such fluctuations at maximum light would also be larger for shorter wavelengths light curves.}

\begin{deluxetable*}{lcccrrrrcccccc}
  \tabletypesize{\scriptsize}
  \tablecaption{Observed Properties of the Miras in Globular Clusters Studied in This Work.\label{tab1}}
  \tablewidth{0pt}
  \tablehead{
    \colhead{Mira} &
    \colhead{$P_l$\tablenotemark{a} (days)} &
    \colhead{$P_n$ (days)} &
    \colhead{$P_c$ (days)} &
    \colhead{$N_g$} &
    \colhead{$N_r$} &
    \colhead{$N_c$\tablenotemark{b}} &
    \colhead{$N_X$\tablenotemark{b}} &
    \colhead{$g_X$} &
    \colhead{$\sigma_g$} &
    \colhead{$r_X$} &
    \colhead{$\sigma_r$} &
    \colhead{$D$ (kpc)\tablenotemark{c}} &
    \colhead{$E$\tablenotemark{d}} 
  }
  \startdata
NGC6356 V1	& 230.6 & 226.7 & 227.0 & 83	 & 423	 & 61	 & 3	 & 15.377 & 0.353 & 13.941 & 0.397 & $15.66\pm0.92$ & $0.366\pm0.002$ \\
NGC6356 V3	& 220.0 & 220.4 & 220.4 & 229	 & 732	 & 100	 & 3	 & 14.778 & 0.397 & 13.426 & 0.487 & $15.66\pm0.92$ & $0.366\pm0.002$ \\
NGC6356 V5	& 219.8 & 220.6 & 220.0 & 256	 & 734	 & 100	 & 4	 & 15.033 & 0.298 & 13.650 & 0.308 & $15.66\pm0.92$ & $0.412\pm0.002$ \\
NGC6638 V3	& 260.0 & 259.5 & 260.8 & 105	 & 584	 & 79	 & 2	 & 15.682 & 0.250 & 13.963 & 0.225 & $9.78\pm0.34$ & $0.410\pm0.003$ \\
NGC6638 V15	& 156.0 & 274.8 & 274.7 & 93	 & 608	 & 73	 & 2	 & 14.512 & 0.197 & 13.052 & 0.065 & $9.78\pm0.34$ & $0.376\pm0.002$ \\
NGC6638 V63	& 213.0 & 214.2 & 215.8 & 122	 & 615	 & 86	 & 3	 & 14.232 & 0.299 & 12.827 & 0.450 & $9.78\pm0.34$ & $0.472\pm0.004$ \\
NGC6642 V1	& 127.0 & 131.1 & 204.6 & 125	 & 621	 & 79	 & 2	 & 14.033 & 0.217 & 12.379 & 0.194 & $8.05\pm0.20$ & $0.484\pm0.002$ \\
NGC6712 V7	& 193.0 & 190.2 & 190.4 & 265	 & 1093	 & 198	 & 4	 & 12.768 & 0.293 & 11.308 & 0.421 & $7.38\pm0.24$ & $0.400\pm0.003$ \\
NGC6760 V3	& 251.0 & 248.4 & 248.7 & 422	 & 2152	 & 411	 & 5	 & 15.193 & 0.261 & 13.407 & 0.273 & $8.41\pm0.43$ & $0.772\pm0.004$ \\
NGC6760 V4	& 226.0 & 225.9 & 226.0 & 642	 & 2770	 & 605	 & 7	 & 15.344 & 0.404 & 13.460 & 0.433 & $8.41\pm0.43$ & $0.772\pm0.004$ \\
Palomar4 V1	& 130.0 & 129.7 & 129.7 & 575	 & 826	 & 451	 & 10	 & 17.438 & 0.264 & 16.332 & 0.337 & $101.39\pm2.57$ & $0.124\pm0.009$ \\
Palomar4 V2	& 150.0 & 150.7 & 150.3 & 557	 & 827	 & 444	 & 8	 & 17.111 & 0.228 & 15.950 & 0.288 & $101.39\pm2.57$ & $0.124\pm0.009$ \\
Palomar7 V1	& 222.0 & 236.4 & 236.6 & 230	 & 688	 & 179	 & 3	 & 16.180 & 0.374 & 14.035 & 0.348 & $4.55\pm0.25$ & $1.142\pm0.002$ \\
Palomar7 V3	& 300.0 & 297.9 & 295.2 & 174	 & 687	 & 147	 & 4	 & 16.006 & 0.499 & 13.814 & 0.387 & $4.55\pm0.25$ & $1.146\pm0.004$ \\
  \enddata
  \tablenotetext{a}{Published period as given in the literature.}
  \tablenotetext{b}{$N_c$ is the number of data-points for constructing the color curves (i.e., pairs of $gr$-band photometry separated within $0.02P_n$~days); $N_X$ is the number of measured magnitudes at maximum light (see text for more details).}
  \tablenotetext{c}{Distance of the host globular clusters adopted from \citet{baumgardt2021}.}
  \tablenotetext{d}{Reddening value returned from the {\tt Bayerstar2019} 3D reddening map \citep{green2019} at the location of the Miras in the globular clusters.}
\end{deluxetable*}

\begin{figure}
  \epsscale{1.1}
  \plotone{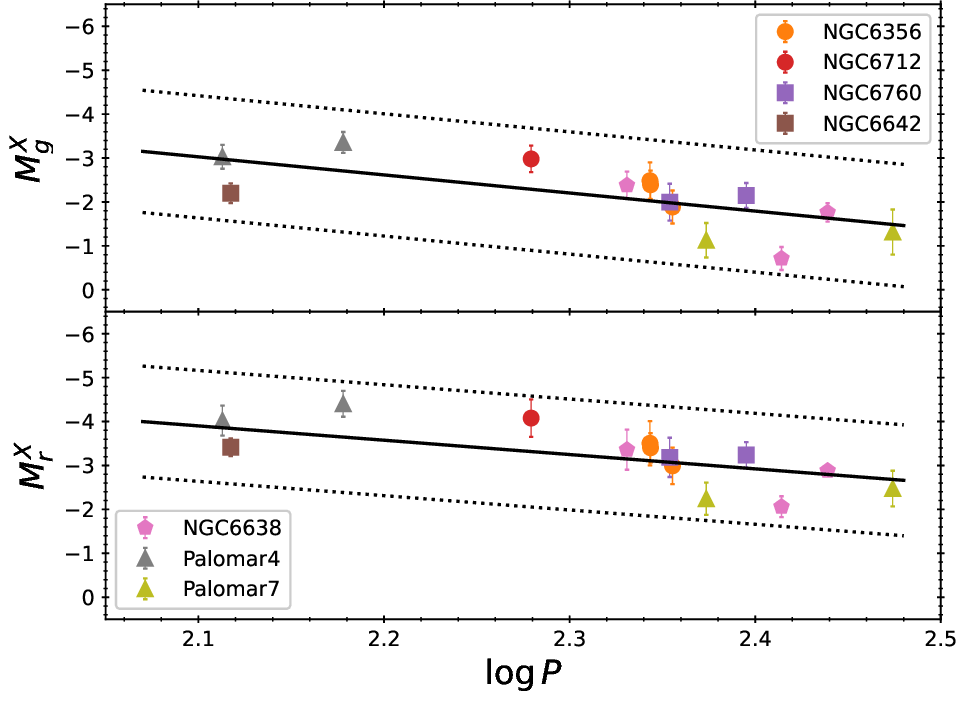}
  \caption{The extinction-corrected PL relations in the $gr$-band for the 14 globular clusters Miras as listed in Table \ref{tab1}. The solid lines  are the fitted PL relations, while the dotted lines represent the $\pm 2.5\sigma$ boundaries. Note that these PL relations are fitted for magnitudes at maximum light, and not the mean lights.}
  \label{fig_plw}
\end{figure}

By adopting the globular clusters distance $D$ (in kpc) from \citet{baumgardt2021} and the reddenings $E$ from the {\tt Bayerstar2019} 3D reddening map \citep[][hence the extinction corrections in the $gr$-band are $3.518E$ and $2.617E$, respectively]{green2019}, we converted the magnitudes at maximum light listed in Table \ref{tab1} to absolute magnitudes ($M^X_m$). The fitted linear PL relations for our sample of 14 globular clusters Miras, as shown in Figure \ref{fig_plw}, are:

\begin{eqnarray}
  M^X_g & = & 4.118 [\pm 0.649](\log P - 2.3) -2.204 [\pm 0.079], \\
  M^X_r & = & 3.262 [\pm 0.562](\log P - 2.3) -3.249 [\pm 0.074],
\end{eqnarray}

\noindent with a dispersion of 0.577~mag and 0.505~mag, respectively. \citet{iwanek2021b} demonstrated that, based on synthetic PL relations at mean light, for wavelengths shorter than $\sim 0.8\mu$m, slopes of the PL relations are positive and decrease with wavelengths, at the same time the zero-points (or intercepts) of the PL relations and the PL dispersions are getting brighter and smaller, respectively. Our derived $gr$-band PL relations follow these trends, even though equation (2) and (3) were derived for magnitudes at maximum light. The positive PL slopes implying that the longer period Miras are fainter than their short period counterparts.

Miras can be classified into oxygen-rich (O-rich) and carbon-rich (C-rich) Miras. Using the determined periods $P_n$ and $JK$-band photometry, adopted either from \citet{sloan2010} or the Two Micron All Sky Survey \citep[2MASS,][]{skrutskie2006}, we classified the 14 globular clusters Miras as O-rich Miras based on the K-nearest neighbor (KNN) algorithm \citep[as done in][]{Ou2023}. Furthermore, for samples of Miras that extended to $\sim 1000$~days the PL relations were often fitted with either a segmented or a quadratic relation \citep[for examples, see][]{ita2011,yuan2017,yuan2018,bhardwaj2019,iwanek2021,ou2022,sanders2023}. In contrast, the longest period in our sample of Miras is $\sim300$~days (Palomar7 V3). Hence, strictly speaking the PL relations we derived in equation (2) \& (3) are applicable to O-rich Miras with period less than 300~days.

\section{Applications to Local Galaxies} \label{sec5}

\begin{deluxetable*}{lccrrrrccccccc}
  \tabletypesize{\scriptsize}
  \tablecaption{Observed Properties and the Derived Distance Modulus for Miras in the Local Galaxies.\label{tab2}}
  \tablewidth{0pt}
  \tablehead{
    \colhead{Mira} &
    \colhead{$P_l$ (days)} &
    \colhead{$P_n$ (days)} &
    \colhead{$N_g$} &
    \colhead{$N_r$} &
    \colhead{$N_c$} &
    \colhead{$N_X$} &
    \colhead{$g_X$} &
    \colhead{$\sigma_g$} &
    \colhead{$r_X$} &
    \colhead{$\sigma_r$} &
    \colhead{$E$\tablenotemark{a}} &
    \colhead{$\mu_g$} &
    \colhead{$\mu_r$}
  }
  \startdata
  \multicolumn{14}{c}{Short-period O-rich Miras} \\
  \hline
  Sextans SDSSJ101234.29-013440.8 & 122 & 120.4 & 369 & 494 & 285&	6 &	16.929 &  0.115	& 15.887 & 0.073 & $0.072\pm0.005$ & $19.78\pm0.67$ & $19.66\pm0.61$ \\
  LeoII CRTSJ111320.6+221116      & 184 & 181.7 & 199 & 454 & 154 &	6 &	19.564 &  0.092	& 18.058 & 0.064 & $0.208\pm0.002$ & $21.20\pm0.59$ & $20.90\pm0.53$ \\
  LeoI L8026	                  & 180 & 198.9 & 88	 & 542 & 78  &	2 &	20.285 &  0.050	& 18.571 & 0.093 & $0.248\pm0.014$ & $21.62\pm0.58$ & $21.18\pm0.53$ \\
  NGC6822 N12557                  & 158 &$\cdots$&$\cdots$& 61&$\cdots$&$\cdots$&$\cdots$&$\cdots$&20.227& 0.032 & $0.316\pm0.002$ & $\cdots$& $23.12\pm0.55$ \\
  NGC6822 N12790                  & 182 & $\cdots$&$\cdots$&42&$\cdots$&$\cdots$&$\cdots$&$\cdots$&20.233& 0.111 & $0.258\pm0.007$ & $\cdots$& $23.08\pm0.54$ \\
  NGC6822 N20540                  & 223 &$\cdots$&$\cdots$&165&$\cdots$&$\cdots$&$\cdots$&$\cdots$&20.200&0.043 & $0.276\pm0.007$ & $\cdots$& $22.72\pm0.53$ \\
  IC1613 G4237                    & 178 &$\cdots$&$\cdots$& 29&$\cdots$&$\cdots$&$\cdots$&$\cdots$&20.991& 0.083 & $0.010\pm0.003$ & $\cdots$& $24.38\pm0.54$ \\
  \hline
  \multicolumn{14}{c}{Long-period O-rich Miras} \\
  \hline
  NGC6822 N20331                  & 314 &$\cdots$&  39 & 149&$\cdots$&$\cdots$& 19.740 & 0.076 & 18.634 & 0.070 & $0.262\pm0.002$ & $\cdots$  & $\cdots$ \\
  NGC6822 N10184                  & 370 &$\cdots$&$\cdots$&119&$\cdots$&$\cdots$&$\cdots$&$\cdots$&19.524 & 0.162&$0.258\pm0.007$ & $\cdots$  & $\cdots$ \\
  NGC6822 N30133                  & 401 &$\cdots$&$\cdots$& 28&$\cdots$&$\cdots$&$\cdots$&$\cdots$&19.243 & 0.021&$0.292\pm0.006$ & $\cdots$  & $\cdots$ \\
  NGC6822 N20134                  & 402 &$\cdots$& 4 & 88&$\cdots$&$\cdots$&$\cdots$&$\cdots$&19.618 & 0.065&$0.270\pm0.004$ & $\cdots$  & $\cdots$ \\
  NGC6822 N40139                  & 545 &$\cdots$& 44&127&$\cdots$&$\cdots$& 20.161 & 0.181 & 18.497 & 0.089&$0.276\pm0.007$ & $\cdots$  & $\cdots$ \\
  NGC6822 N10198                  & 602 &$\cdots$& 30&244&$\cdots$&$\cdots$& 20.400 & 0.069 & 19.462 & 0.018&$0.206\pm0.004$ & $\cdots$  & $\cdots$ \\
  NGC6822 N30292                  & 637 &$\cdots$&  7&203&$\cdots$&$\cdots$&$\cdots$&$\cdots$&19.470 & 0.161&$0.302\pm0.004$ & $\cdots$  & $\cdots$ \\
  NGC6822 N10091                  & 638 &$\cdots$&  3& 39&$\cdots$&$\cdots$&$\cdots$&$\cdots$&18.918 & 0.028&$0.206\pm0.004$ & $\cdots$  & $\cdots$ \\
  NGC6822 N20004                  & 854 &693.6&180&396& 168	  & 2	   & 18.033 & 0.327  &16.351 & 0.363&$0.276\pm0.007$ & $\cdots$  & $\cdots$ \\
  IC1613 G1016                    & 464 &$\cdots$& 141 &397 &$\cdots$&$\cdots$& 20.021 & 0.092 & 18.825 & 0.088 & $0.000\pm0.000$ & $\cdots$  & $\cdots$ \\
  IC1613 G1017                    & 580 &$\cdots$&  83 &191 &$\cdots$&$\cdots$& 19.605 & 0.106 & 18.508 & 0.100 & $0.000\pm0.000$ & $\cdots$  & $\cdots$ \\
  IC1613 G2035                    & 530 &$\cdots$&  65 &212 &$\cdots$&$\cdots$& 19.832 & 0.081 & 18.730 & 0.059 & $0.000\pm0.000$ & $\cdots$  & $\cdots$ \\
  IC1613 G3011                    & 550 &$\cdots$&  42 &137 &$\cdots$&$\cdots$& 19.606 & 0.041 & 18.581 & 0.057 & $0.000\pm0.000$ & $\cdots$  & $\cdots$ \\
  \hline
  \multicolumn{14}{c}{C-rich Miras} \\
  \hline
  LeoI L1019	                  & 158 &160.5&  54 &470 &41	    &1&	20.634&$\cdots$&18.981&$\cdots$& $0.248\pm0.014$ & $22.36\pm0.60$ & $21.89\pm0.54$ \\
  LeoI L1077                      & 336 &$\cdots$&$\cdots$&55&$\cdots$&$\cdots$&$\cdots$&$\cdots$& 20.130 & 0.217 &$0.178\pm0.007$ & $\cdots$  & $22.17\pm0.62$ \\
  NGC6822 N10817                  & 214 &$\cdots$& 100    &327&$\cdots$&$\cdots$& 20.238 & 0.089 & 19.831 & 0.054 &$0.280\pm0.005$ & $21.33\pm0.58$ &  $22.25\pm0.53$ \\
  NGC6822 N40590                  & 223 &$\cdots$&$\cdots$& 78&$\cdots$&$\cdots$&$\cdots$&$\cdots$& 20.946 & 0.032&$0.206\pm0.004$ & $\cdots$  & $23.50\pm0.52$ \\
  NGC6822 N12751                  & 231 &$\cdots$&$\cdots$&140&$\cdots$&$\cdots$&$\cdots$&$\cdots$& 20.389 & 0.098&$0.258\pm0.007$ & $\cdots$  & $22.76\pm0.53$ \\
  NGC6822 N11032                  & 239 &$\cdots$&$\cdots$&104&$\cdots$&$\cdots$&$\cdots$&$\cdots$& 20.703 & 0.046&$0.316\pm0.002$ & $\cdots$  & $22.87\pm0.53$ \\
  NGC6822 N20578                  & 246 &$\cdots$&$\cdots$& 61&$\cdots$&$\cdots$&$\cdots$&$\cdots$& 20.437 & 0.084&$0.236\pm0.002$ & $\cdots$  & $22.77\pm0.54$ \\
  NGC6822 N20542                  & 255 &$\cdots$&$\cdots$&190&$\cdots$&$\cdots$&$\cdots$&$\cdots$& 20.251 & 0.034&$0.276\pm0.007$ & $\cdots$  & $22.43\pm0.53$ \\
  NGC6822 N30430                  & 269 &$\cdots$&$\cdots$& 59&$\cdots$&$\cdots$&$\cdots$&$\cdots$& 20.405 & 0.080&$0.292\pm0.006$ & $\cdots$  & $22.47\pm0.55$ \\
  NGC6822 N12208                  & 278 &$\cdots$& 12     & 92&$\cdots$&$\cdots$&$\cdots$&$\cdots$& 19.879 & 0.022&$0.280\pm0.005$ & $\cdots$  & $21.93\pm0.55$ \\
  NGC6822 N30583                  & 302  &$\cdots$& 1     & 76&$\cdots$&$\cdots$&$\cdots$&$\cdots$& 20.954 & 0.074&$0.254\pm0.002$ & $\cdots$  & $22.95\pm0.56$ \\
  NGC6822 N40114                  & 312  &$\cdots$& 4     & 35&$\cdots$&$\cdots$&$\cdots$&$\cdots$& 20.835 & 0.062&$0.302\pm0.004$ & $\cdots$  & $22.66\pm0.57$ \\
  NGC6822 N11059                  & 319  &$\cdots$& 1     & 25&$\cdots$&$\cdots$&$\cdots$&$\cdots$& 20.728 & 0.259&$0.258\pm0.007$ & $\cdots$  & $22.64\pm0.63$ \\
  NGC6822 N20657                  & 343 &$\cdots$&$\cdots$& 25&$\cdots$&$\cdots$&$\cdots$&$\cdots$& 20.705 & 0.183&$0.262\pm0.002$ & $\cdots$  & $22.50\pm0.61$ \\
  NGC6822 N40520                  & 432 &$\cdots$& 3     & 34&$\cdots$&$\cdots$&$\cdots$&$\cdots$& 20.791  & 0.152&$0.206\pm0.004$ & $\cdots$  & $22.41\pm0.67$ \\
  NGC6822 N10753                  & 432 &$\cdots$&$\cdots$&115&$\cdots$&$\cdots$&$\cdots$&$\cdots$& 20.700 & 0.109&$0.280\pm0.005$ & $\cdots$  & $22.12\pm0.66$ \\
  NGC6822 N21141                  & 456 &$\cdots$&$\cdots$&107&$\cdots$&$\cdots$&$\cdots$&$\cdots$& 19.894 & 0.068&$0.236\pm0.002$ & $\cdots$  & $21.36\pm0.67$ \\
  IC1613 G4251                    & 263 &$\cdots$&$\cdots$& 34&$\cdots$&$\cdots$&$\cdots$&$\cdots$& 21.050 & 0.068 &$0.000\pm0.000$ & $\cdots$  & $23.91\pm0.54$ \\
  IC1613 G3083                    & 280 &$\cdots$&$\cdots$& 42&$\cdots$&$\cdots$&$\cdots$&$\cdots$& 20.689 & 0.229 &$0.000\pm0.000$ & $\cdots$  & $23.46\pm0.59$ \\
  IC1613 G3144                    & 364 &$\cdots$&$\cdots$& 31&$\cdots$&$\cdots$&$\cdots$&$\cdots$& 20.671 & 0.175 &$0.000\pm0.000$ & $\cdots$  & $23.07\pm0.63$ \\
  IC1613 G4183                    & 410 &$\cdots$&$\cdots$& 17&$\cdots$&$\cdots$&$\cdots$&$\cdots$& 21.040 & 0.066 &$0.000\pm0.000$ & $\cdots$  & $23.27\pm0.64$ \\
  \enddata
  \tablenotetext{a}{Reddening value returned from the {\tt Bayerstar2019} 3D reddening map \citep{green2019}.}
\end{deluxetable*}

\citet{Ou2023} demonstrated that ZTF could detect the peak brightness of known Miras in M33, which is located at a distance of 0.86~Mpc \citep[or $\mu=24.67\pm0.07$~mag,][]{dgb2014}. Therefore, we searched for local galaxies that are closer than M33, located within the visibility of ZTF, and contain known Miras detected from NIR observations. We identified five such galaxies: Sextans \citep{sakamoto2012}, Leo~I \citep{menzies2002,menzies2010}, Leo~II \citep{grady2019}, NGC6822 \citep{whitelock2013b}, and IC1613 \citep{menzies2015}.  Literature distance modulus $\mu$ for Leo~I, Leo~II, and Sextans were adopted from \citet{dw2020} as $\mu_{\mathrm{Sextans}}=19.7$~mag, $\mu_{\mathrm{Leo~I}}=22.0$~mag, and $\mu_{\mathrm{Leo~II}}=21.8$~mag, respectively. In case of NGC6822 and IC1613, \citet{parada2023} summarized recent distance measurements to these two galaxies. We adopted the mid-point of the maximum and minimum $\mu$ listed in their Table 10 and 11, as $\mu_{\mathrm{NGC6822}}=23.4$~mag and $\mu_{\mathrm{IC1613}}=24.3$~mag \citep[this value is also same as the recommended value given in][]{dgb2014}, respectively.

\begin{figure}
  \epsscale{1.1}
  \plotone{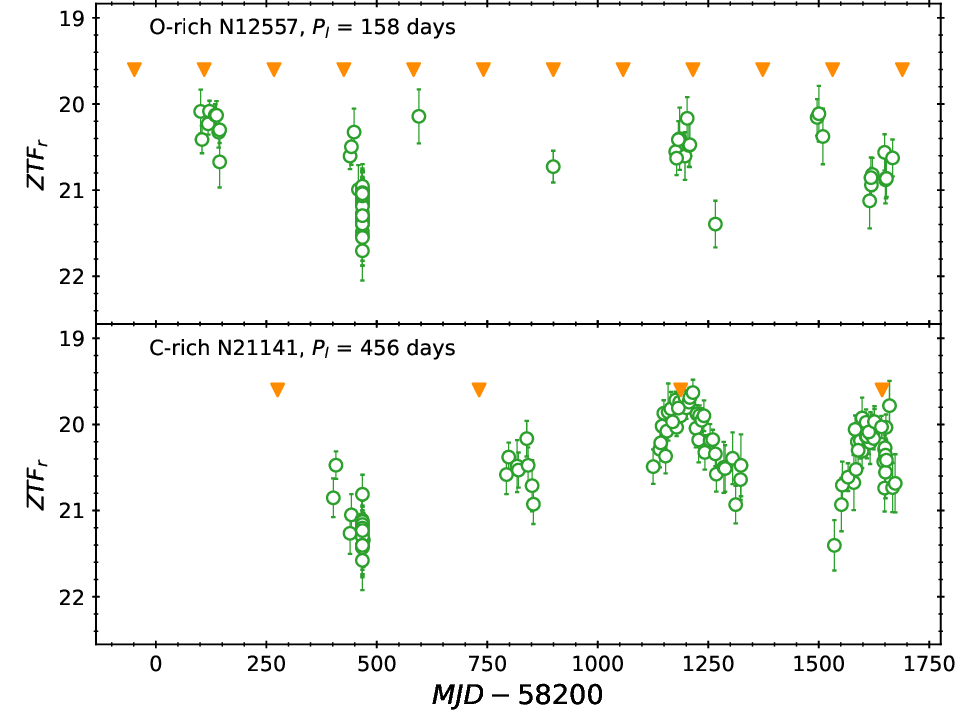}
  \caption{Examples of $r$-band ZTF light curves for two Miras in NGC6822. The orange triangles marked the expected epochs of maximum light (based on the literature period), and the brightest observed data-points tend to be found near these orange triangles.}
  \label{fig_portionlc}
\end{figure}

Three extragalactic Miras (L2077 in Leo~I, N21029 in NGC6822, and SDSSJ101525.93-020431.8 in Sextans) are known to exhibit a (periodic) long-term trend, and hence they were excluded from the sample. For the rest of the Miras in these five local galaxies, we extracted their light curves from the ZTF DR16 and the partner survey data (in the same way as the globular clusters Miras, see Section \ref{sec2}). Nevertheless, about half of them either do not have ZTF light curve data or only contain very few data points in the $g$- and/or $r$-band (mostly for the two distant galaxies NGC6822 and IC1613). For the remaining Miras, we re-classified them into the O-rich and C-rich Miras based on their $JK$-band photometry published in the aforementioned publications and the same KNN algorithms applied on the globular clusters Miras.

Based on the quality and the number of data-points on the ZTF light curves, we analyzed these Miras using two approaches. In case there are enough number of data-points in the $gr$-band ZTF light curves, we followed the same procedures as the globular clusters Miras to determine their $P_n$, $m_X$ and $\sigma_X$. This approach was only applied to five Miras. Otherwise we adopted the literature periods and only determine the $m_X$ and $\sigma_X$, mostly in the $r$-band, by taking the means of the brightest 10\% data-points \citep[that is, following the approach as outlined in][]{bhardwaj2019}. This is because for Miras in the distant galaxies, it is possible that only the portion of the light curves around maximum light could be detected by ZTF \citep[note that the limiting magnitude for ZTF is $r\sim 20.6$~mag,][]{bellm2019}, as illustrated in Figure \ref{fig_portionlc}. We then applied the color-term corrections on these brightest 10\% data-points by adding $C_m(g_X-r_X)$ to their calibrated magnitudes, where $C_m$ were extracted from the ZTF PSF catalogs for individual data-points. Assuming the $(g-r)$ colors at maximum light can be approximated with $(g_X-r_X)$, and fitting the globular clusters Miras in Table \ref{tab1} yields:

\begin{eqnarray}
  (g_X-r_X)\ & = &\  0.910[\pm1.763] (\log P - 2.3) \nonumber \\
            &  & + 1.522[\pm0.216],
\end{eqnarray}

\noindent with a dispersion of $0.299$~mag.\footnote{The extinction-corrected version of this relation is: $(g_X-r_X)_0 = 0.110[\pm0.846] (\log P - 2.3) + 1.124[\pm0.107]$, with a dispersion of $0.103$~mag.} The measured $P_n$ (if applicable), $m_X$, $\sigma_X$, and other relevant information are summarized in Table \ref{tab2}.

\begin{figure*}
  \centering
  \begin{tabular}{ccc}
    \includegraphics[width=0.32\textwidth]{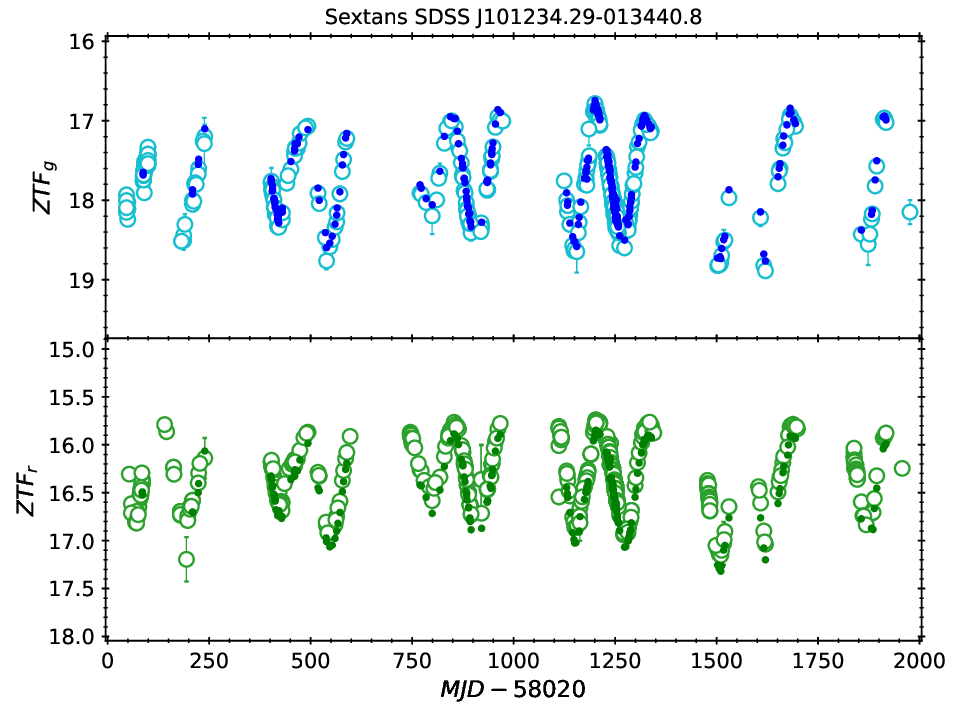} &  \includegraphics[width=0.32\textwidth]{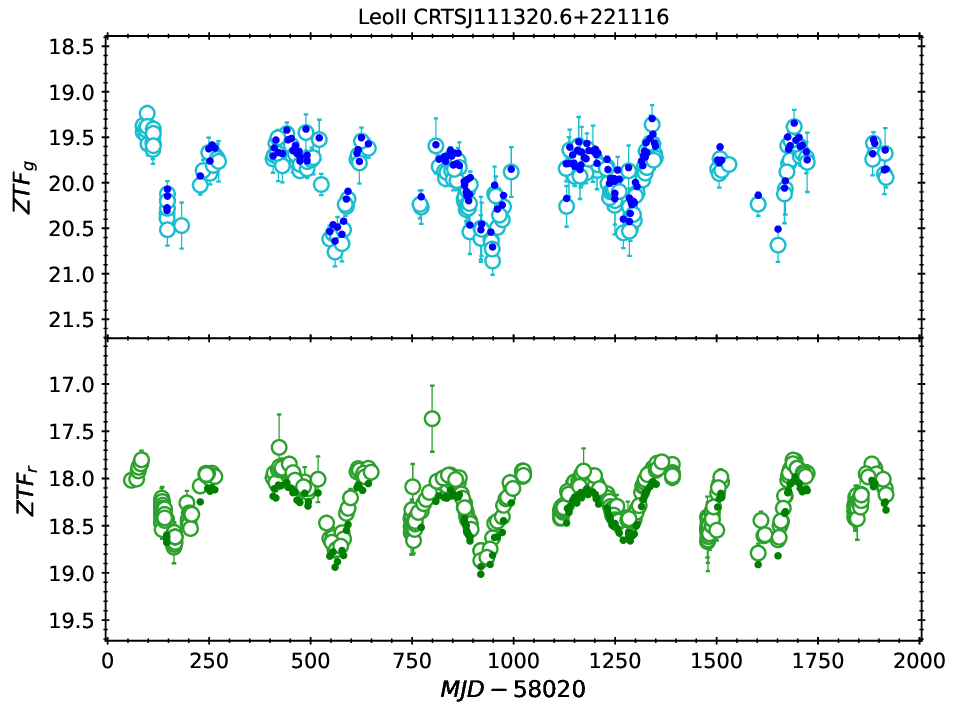} & \includegraphics[width=0.32\textwidth]{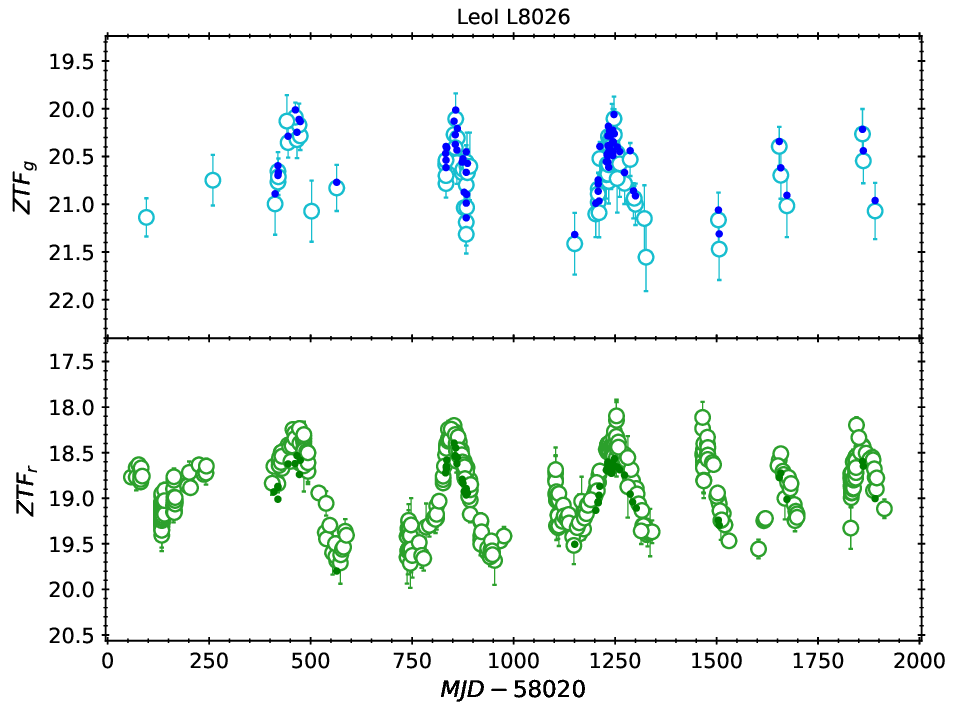} \\
  \end{tabular}
\caption{The $gr$-band ZTF light curves for the three short-period O-rich extragalactic Miras. The symbols are same as in the right panels of Figure \ref{fig_lcc}. }
\label{fig_oshort}
\end{figure*}

\subsection{Short-Period O-Rich Miras}\label{so}

\begin{figure*}
  \epsscale{1.1}
  \plottwo{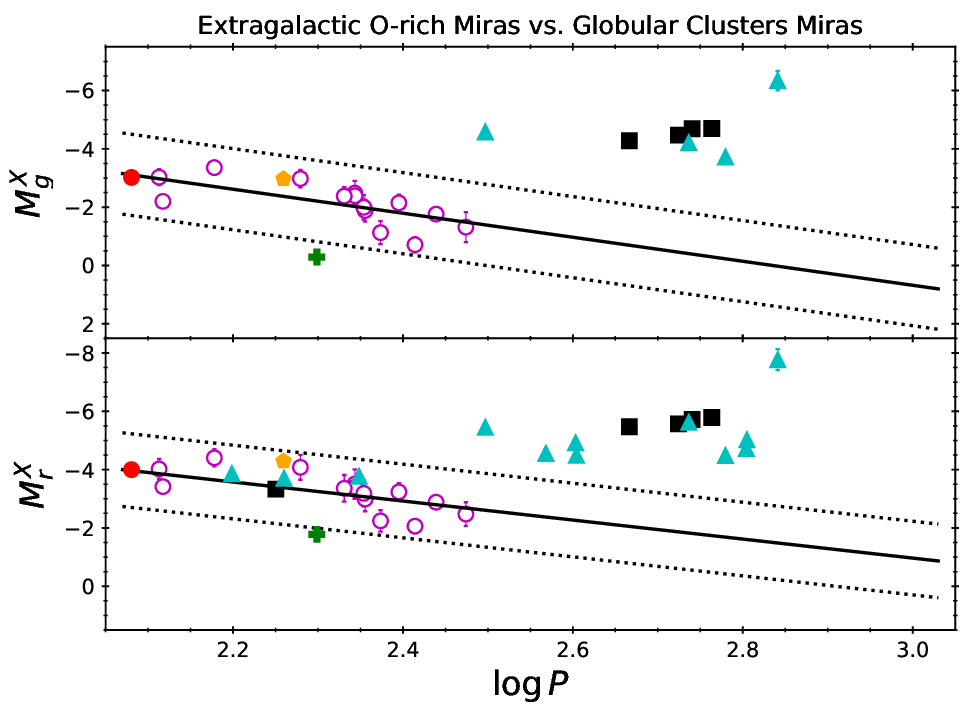}{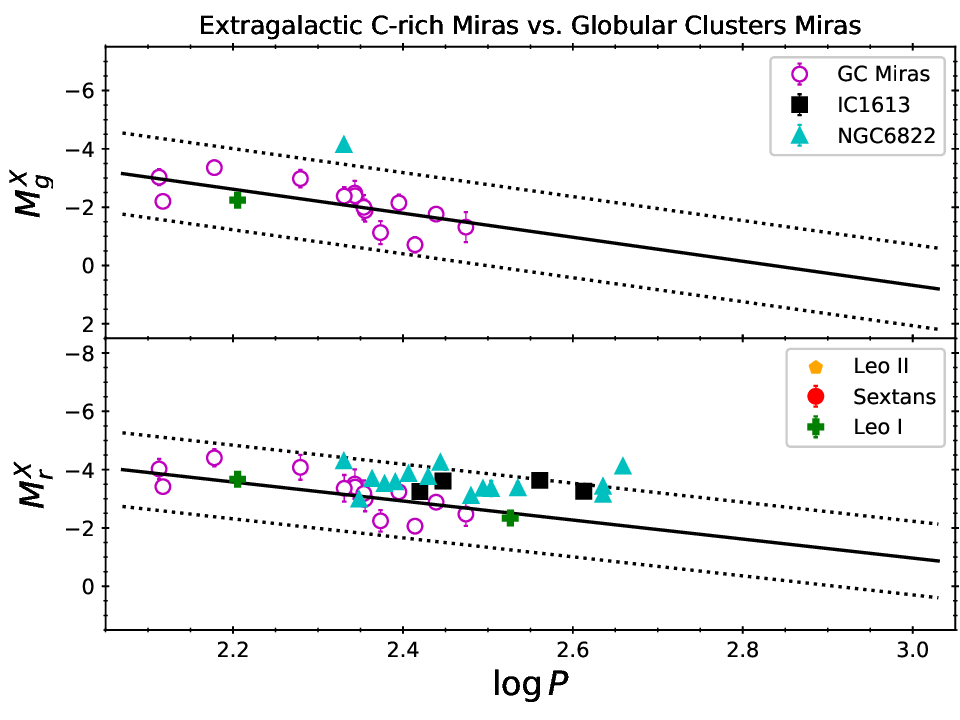}
  \caption{Comparison of the $gr$-band absolute magnitudes at maximum light between globular clusters Miras (open magenta circles) and extragalactic Miras (in various symbols). For extragalactic Miras, literature $\mu$ (see text for details) were adopted to convert $m_X$ to $M^X_m$ (after corrected for extinction). The left and right panels are for the comparisons of O-rich Miras and C-rich Miras, respectively. The lines are same as in Figure \ref{fig_plw}, but extrapolated to a longer period.}
\label{fig_plcompare}
\end{figure*}  

There are seven O-rich Miras with period shorter than 300~days (i.e. short period) detected in all five galaxies. Figure \ref{fig_oshort} presents three Miras which have ZTF light curves in both $gr$-band. We calculated the distance moduli for these seven Miras using our derived PL relations (presented in Section \ref{sec4}), they are listed in the last two columns of Table \ref{tab2}. The derived distance moduli for Sextans SDSSJ101234.29-013440.8, NGC6822 N12557 and N12790, as well as IC1613 G4237 are in good agreement with the literature values. The Mira CRTSJ111320.6+221116 in Leo~II has a smaller distance modulus, especially in the $r$-band, when compared to the literature $\mu_{\mathrm{Leo~II}}=21.8$~mag. \citet{grady2019} also found a shorter distance ($193\pm15$~kpc) for this Mira, using a totally independent dataset and PL relation, with respect to Leo~II ($233\pm14$~kpc). Therefore, our result is consistent with the finding of \citet{grady2019}. The $r$-band distance moduli for Leo~I L8026 and NGC6822 N20540 were also smaller, nevertheless they are still fall within $\sim \pm1.5\sigma$ from the literature values.

By adopting the literature $\mu$, the absolute magnitudes of these seven short-period extragalactic O-rich Miras were compared to the globular clusters Miras in the left panels of Figure \ref{fig_plcompare}. With the exception of Leo~I L8026, all other short-period O-rich Miras are consistent with globular cluster Miras and located within the $\pm2.5\sigma$ boundaries of the $r$-band PL relation.

\subsection{Long-Period O-Rich Miras}\label{lo}

NGC6822 and IC1613 are the only two galaxies in our sample containing O-rich Miras with period longer than 300~days. The long-period O-rich Miras in NGC6822 and IC1613, as shown in \citet{whitelock2013b} and \citet{menzies2015}, respectively, were known to be over-luminous in the $K$-band and the bolometric magnitude. These over-luminous Miras were presumably undergoing hot bottom burning (HBB) phase \citep[for examples, see][]{whitelock2003,ita2011}. For the long-period O-rich Miras listed in Table \ref{tab2}, their $gr$-band absolute magnitudes were also much brighter than the magnitudes predicted from extrapolating the PL relations to a longer period, as demonstrated in the left panels of Figure \ref{fig_plcompare}. Interestingly, after excluding the longest-period Mira, these long-period O-rich Miras seems to have a near constant or mildly period-dependent $gr$-band absolute magnitudes at maximum light. Since the ZTF light curves are incomplete for these long-period Miras, and color-term corrections using equation (4) might not be valid for them, the existence of such a near constant absolute magnitude is not conclusive. The much deeper and 10~years light curves data collected from LSST can be used to investigate these long-period Miras further, because both galaxies are located within the footprint of LSST.

\subsection{C-Rich Miras}\label{cmira}

In the era of synoptic sky surveys such as LSST, newly discovered extragalactic Miras may lack NIR photometry or spectroscopic observations to be classified to O-rich or C-rich. Hence, the C-rich Miras in Leo~I, NGC6822 and IC1613 provide an opportunity to test the distance measurement if a genuine C-rich Miras discovered from optical surveys was mis-identified as an O-rich Mira. However, the majority of them only contain incomplete $r$-band light curves, and suffer the same problems as the long-period O-rich Miras (see the discussion in sub-section \ref{lo}), the derived distance moduli for these C-rich Miras, listed in the last two columns of Table \ref{tab2} using equation (2) and (3), should be treated with caution.

The $gr$-band distance moduli of the two C-rich Miras in Leo~I are in good agreement with the literature $\mu$ of $\mu_{\mathrm{Leo~I}}=22.0$~mag. In the case of C-rich Miras in both NGC6822 and IC1613, except for a few short-period Miras, the derived distance moduli were smaller than the literature values. In other words, they seems to be slightly over-luminous at a given period, as shown in the right panels of Figure \ref{fig_plcompare}. The long-period C-rich Miras are also tend to be over-luminous, similar to the cases of long-period O-rich Miras. The right panels of Figure \ref{fig_plcompare} suggested the C-rich Miras seems to have a flat PL relation in the $r$-band. 

\section{Discussion and Conclusions} \label{sec6}

In this work, we derived the $gr$-band PL relations for the O-rich and short-period ($<300$~days) Miras in the globular clusters using the light-curve data collected from ZTF, as these globular clusters possess homogeneous distances compiled in \citet{baumgardt2021}. We focus on the maximum light when deriving the PL relations because Miras are known to exhibit a smaller PL dispersion at maximum light when compared to their mean-light counterparts. Furthermore, the $(g-r)$ colors tend to be bluest at maximum light, implying the color-term corrections, $C_m(g-r)$, would be smaller when compared to other pulsation phases such as at mean light. Finally, due to the large-amplitude nature of Miras, it is possible that only the portions of the light curves around the maximum light could be detected for Miras located in distant galaxies. In this scenario, measurements at maximum light can still be used to derive the distance moduli to the host galaxies, as we demonstrated in Section \ref{sec5} for the Miras in NGC6822 and IC1613. Given that the limiting magnitude for ZTF is $r\sim 20.6$~mag \citep{bellm2019}, the maximum light for a Mira with $\log P=2.3$, based on equation (3), could be detected at a distance modulus of $\mu\sim23.9$~mag, which is within $\pm0.5$~mag from $\mu_{\mathrm{NGC6822}}=23.4$~mag and $\mu_{\mathrm{IC1613}}=24.3$~mag. For LSST, the single-epoch limiting magnitude in the $r$-band is $\sim24.7$~mag, implying a $\log P=2.3$ Mira could be detected up to $\mu\sim 28$~mag at maximum light, corresponding to a distance of $\sim 4$~Mpc.

\begin{figure}
  \epsscale{1.1}
  \plotone{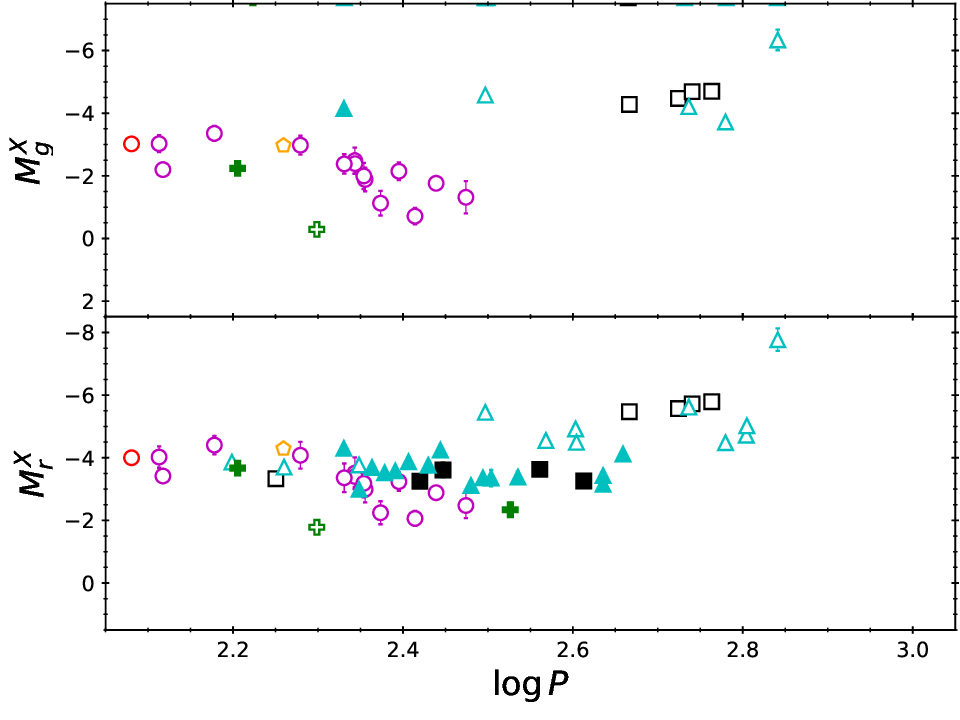}
  \caption{The PL relations at maximum lights for all of the Miras listed in Table \ref{tab1} and \ref{tab2}. The symbols are same as in Figure \ref{fig_plcompare}. However, for a better visualization, open and filled symbols were used represent all of the O-rich and C-rich Miras, respectively.}
  \label{fig_allpl}
\end{figure}

In Figure \ref{fig_allpl}, we presented the $gr$-band PL relations at maximum light for all of the Miras investigated in this work. In the $r$-band, which has more Miras than the $g$-band, both of the short-period O-rich and C-rich Miras seem to occupy the same region in the PL plane, suggesting our derived $gr$-band PL relations could also be used for the short-period C-rich Miras (also, see the discussion in Section \ref{cmira}). However, as periods become longer, the long-period C-rich Miras appear to ``connect'' to the long-period HBB O-rich Miras. Altogether, the short- and long-period Miras formed a quadratic or segmented relation, hence the linear PL relations should only be applied to the short-period Miras. We suggest using short-period extragalactic Miras detected with LSST for distance measurements not only because they follow a linear relation, but also their periods will be better constraint from the 10~years observations.

\acknowledgments

We are thankful for funding from the National Science and Technology Council (Taiwan) under the contracts 107-2119-M-008-014-MY2, 107-2119-M-008-012, 108-2628-M-007-005-RSP and 109-2112-M-008-014-MY3.

Based on observations obtained with the Samuel Oschin Telescope 48-inch Telescope at the Palomar Observatory as part of the Zwicky Transient Facility project. ZTF is supported by the National Science Foundation under Grants No. AST-1440341 and AST-2034437 and a collaboration including current partners Caltech, IPAC, the Weizmann Institute of Science, the Oskar Klein Center at Stockholm University, the University of Maryland, Deutsches Elektronen-Synchrotron and Humboldt University, the TANGO Consortium of Taiwan, the University of Wisconsin at Milwaukee, Trinity College Dublin, Lawrence Livermore National Laboratories, IN2P3, University of Warwick, Ruhr University Bochum, Northwestern University and former partners the University of Washington, Los Alamos National Laboratories, and Lawrence Berkeley National Laboratories. Operations are conducted by COO, IPAC, and UW.

This publication makes use of data products from the Two Micron All Sky Survey, which is a joint project of the University of Massachusetts and the Infrared Processing and Analysis Center/California Institute of Technology, funded by the National Aeronautics and Space Administration and the National Science Foundation.

This research has made use of the SIMBAD database and the VizieR catalogue access tool, operated at CDS, Strasbourg, France. This research made use of Astropy,\footnote{\url{http://www.astropy.org}} a community-developed core Python package for Astronomy \citep{astropy2013, astropy2018, astropy2022}.

\facility{PO:1.2m}

\software{{\tt astropy} \citep{astropy2013,astropy2018,astropy2022}, {\tt dustmaps} \citep{green2018}, {\tt gatspy} \citep{vdp2015}, {\tt Matplotlib} \citep{hunter2007},  {\tt NumPy} \citep{harris2020}, {\tt SciPy} \citep{virtanen2020}, {\tt statsmodels} \citep{seabold2010}.}



\end{document}